\documentclass[aps,prx,twocolumn,floats,showpacs,superscriptaddress,nofootinbib]{revtex4-1}
\usepackage{graphicx,epsfig}
\usepackage{times,bbm}
\usepackage{graphics,dcolumn,bm,float}
\usepackage{amssymb,amsmath,rotate,color}
\usepackage[title,titletoc,toc]{appendix}
\usepackage{mathtools}
\usepackage{booktabs}
\usepackage{tcolorbox}

\usepackage[pagebackref=false,colorlinks,linkcolor=magenta,citecolor=cyan,urlcolor=magenta]{hyperref}

\begin{document}
	\unitlength 1 cm
	\newcommand{\be}{\begin{equation}}
	\newcommand{\ee}{\end{equation}}
	\newcommand{\nn}{\nonumber}
	\newcommand{\vk}{\vec k}
	\newcommand{\vp}{\vec p}
	\newcommand{\vq}{\vec q}
	\newcommand{\vkp}{\vec {k'}}
	\newcommand{\vpp}{\vec {p'}}
	\newcommand{\vqp}{\vec {q'}}
	\newcommand{\bsq}{{\boldsymbol{q}}}
	\newcommand{\bsk}{{\boldsymbol{k}}}
	\newcommand{\up}{\uparrow}
	\newcommand{\down}{\downarrow}
	\newcommand{\cdag}{c^{\dagger}}
	\newcommand{\hlt}[1]{\textcolor{red}{#1}}
	\newcommand{\ba}{\begin{align}}
	\newcommand{\ea}{\end{align}}
	\newcommand{\la}{\langle}
	\newcommand{\ra}{\rangle}
	\newcommand{\bearr}{\begin{eqnarray}}
	\newcommand{\eearr}{\end{eqnarray}}
	\newcommand{\eps}{\varepsilon}
	\newcommand{\sgn}{\text{sgn}}
	\newcommand{\bq}{{\boldsymbol{q}}}
	\newcommand{\bk}{{\boldsymbol{k}}}
	\newcommand{\bet}{{\boldsymbol{\eta}}}
	\newcommand{\btau}{{\boldsymbol{\tau}}}
	\newcommand{\bE}{{\boldsymbol{E}}}
	\newcommand{\bB}{{\boldsymbol{B}}}
	\newcommand{\bp}{{\boldsymbol{p}}}
	\newcommand{\bv}{{\boldsymbol{v}}}
	\newcommand{\br}{{\boldsymbol{r}}}
	\newcommand{\pr}{\partial}
	\newcommand{\bs}{\boldsymbol}
	\newcommand{\etal}{{\it et al}}
	\newcommand{\bmt}{\left[\begin{matrix}}
		\newcommand{\emt}{\end{matrix}\right]}

	\author{T. Farajollahpour}
	\affiliation{Department of Physics, Sharif University of Technology, Tehran 11155-9161, Iran}

	\author{S. A. Jafari}
	\email{akbar.jafari@gmail.com}
	\affiliation{Department of Physics, Sharif University of Technology, Tehran 11155-9161, Iran}
	\affiliation{Center of excellence for Complex Systems and Condensed Matter (CSCM), Sharif University of Technology, Tehran 1458889694, Iran}
	
	\date{\today}
	
	\title{Synthetic non-Abelian gauge fields and gravitomagnetic effects in tilted Dirac cone systems}
	
	\begin{abstract}
	In planar tilted Dirac cone systems, the tilt parameter can be made space-dependent by either a perpendicular displacement field,
	or by chemical substitution in certain systems. We show that the symmetric partial derivative of the tilt parameter generates non-Abelian 
	synthetic gauge fields in these systems. 
	The small velocity limit of these gauge forces corresponds to Rashba and Dresselhaus spin-orbit couplings. At the classical level,
	the same symmetric spatial derivatives of tilt contribute to conservative, Lorentz-type and friction-like forces. The velocity dependent 
	forces are odd with respect to tilt and therefore have opposite signs in the two valleys when the system is inversion symmetric. 
	Furthermore, toggling the chemical potential between the valence and conduction bands reverses the sign of the all these classical forces,
	which indicates these forces couple to the electric charge of the carriers. As such, these forces are natural extensions of the electric
	and magnetic forces in the particular geometry of the tilted Dirac cone systems. 
	\end{abstract}

	\maketitle

\section{Introduction}
In solid state physics, the lattice breaks the Lorentz symmetry of the vacuum. 
But in certain lattices, such as the honeycomb lattice of graphene, the Lorentz symmetry emerges in a lower energy scale
and with a velocity scale, $v_F$ which is much smaller than the seed of light~\footnote{In general, a mass term $\gamma^0 m$ can also be included.}
\be
    H_D= v_F \gamma^0\gamma^i p_i+mv_F^2\gamma^0,
\ee
where a possibly non-zero Dirac mass $m$ corresponds to the band gap. 
Here $d$ is the space dimensions, $i=1\ldots d$ and $\gamma^\mu$ are Dirac matrices~\cite{ZeeBook,PeskinBook}. 
As usual Greek indices run from $0,\ldots,d$ with $\mu=0$ denoting the "time". 
The Dirac cone in the dispersion relation of the electronic degrees of freedom in Dirac solids at one level deeper can be attributed to an 
emergent effective Minkowski spacetime (at length scales much larger than the lattice constant), from which the Lorentz 
symmetry immediately and quite naturally follows. 

But unlike the Lorentz symmetry of the vacuum of the standard model of particle physics,
the emergent Lorentz symmetry of the condensed matter systems is not stringent and can be broken in 
a number of interesting ways. One way to break the Lorentz symmetry is to {\em tilt} the Dirac/Weyl cone in the spectrum.
This can be done by adding a term $\hbar v_F\zeta^i p_i$ proportional to unit matrix of appropriate dimension to give,
\be
    H_D= v_F \gamma^0\gamma^i p_i+mv_F^2\gamma^0+\hbar v_F\zeta^i p_i,
    \label{TD.eqn}
\ee
There are two approaches to study the Hamiltonian~\eqref{TD.eqn}: (i) The first one is the standard
solid-state approach to take this equation as the starting point and study its consequences and
explore the effect of "tilt parameters" $\bs\zeta\equiv \zeta^i$ in various physical properties. 
The second approach is to start from an effective spacetime structure specified by a metric $g_{\mu\nu}(\zeta^i)$
that reduces to the Minkowski structure as $\zeta^i\to 0$, namely $g_{\mu\nu}(\zeta^i\to 0)=\eta_{\mu\nu}={\rm diag}(-1,\delta_{ij})$
where $\delta_{ij}$ is the Kronecker delta function. 

We start by providing a comprehensive introduction to materials in various dimensions supporting the tilted cone
dispersion, followed by a subsection introducing the spacetime structure of these materials.

\subsection{Materials hosting tilted Dirac/Weyl cones} 
Such tilted Dirac cone materials (TDCMs) and corresponding tilted Dirac fermions (TDFs) can exist
in three (3D), two (2D) and even one (1D) space dimensions. 
For historical reasons, let us start the discussion with 2D TDFs. 
Historically, he layered organic compound $\alpha$-(BEDT-TTF)$_2$I$_3$ was the first realization of tilted Dirac 
cone~\cite{Tajima2006,Katayama2006,Kobayashi2007,Tajima2009,Kobayashi2009,Isobe2017} which was discovered in Japan~\cite{Kajita2014}.
The Dirac structure of charge carriers in this compound is imprinted 
in their $\pi$ Berry phase inferred from their Landau quantization~\cite{Tajima2013QHEtilted}.
The tilt of the Dirac cone in organic compound can be inferred from interlayer magnetoresistance~\cite{Himura2009,Sugawara2010}.
Landau levels were obtained from semiclassical quantization~\cite{Goerbig2008}. Quantum mechanically solution in the presence of an
in plane electric field, it was found that the valley degeneracy is lifted~\cite{Goerbig2009}.
Magnetoplasmons in organic compound was studied by S\'ari \etal~\cite{Goerbig2014b}.
NMR measurements suggest that this compound is a strongly correlated TDF system~\cite{Hirata2011NRM} marked by three orders of 
magnitude enhancement of the Korringa ratio~\cite{Hirata2017} and the reshaping of Dirac cone caused by strong Coulomb interactions in organic TDF system~\cite{Hirata2016}.\\

The list of TDFs in 2D is expanding rapidly. Starting from the above organic compound, substitution of iodine 
with halogens were examined, and it was found that replacement I$\to$F can over-tilt the Dirac cone~\cite{Geilhufe2018}.
This is particularly important as it corresponds to transition from type-I to type-II Dirac fermions. 
Type-II 2D DFs are suggested by {\em ab-initio} calculations for quantum wells of LaAlO$_3$/LaNiO$_3$/LaAlO$_3$~\cite{Tao2018Tilted},
where varying the number of LaNiO$_3$ layers can shift the Dirac node. The surface of 
crystalline topological insulators~\cite{Schnyder2017Tilted} is also predicted to host TDFs. 
Furthermore, the antiferromagnetic phase of the Iron-based superconductors 
as excitations above a spin density wave mean field state~\cite{Tohyama2010IronBasedTilted}. 
Angular resolved photoemission measurements of Varykhalov \etal which is also supported by {\em ab-initio} calculations,
find a tilted Dirac cone on the metallic surface of W caused by Rashba spin-orbit interaction~\cite{Varykhalov2017}.
Ultra-low dissipation in the conductivity of BaFe$_2$As$_2$ was attributed to the tilted Dirac cone in its spectrum~\cite{Imai2013}.

The above examples are 2D TDFs based on layered compounds. TDFs are also expected in purely 
2D, namely one-atom-thick compounds. 
First principle calculations suggest that partially hydrogenated graphene compound, C$_6$H$_2$ has
a tilted Dirac cone (TDC) in its spectrum~\cite{Lu2016C6H2}. The left neighbor of Carbon in the periodic table of elements, namely, 
boron in $8Pmmn$ lattice structure is also predicted to host TDFs~\cite{Zhou2014,Lopez2016}. Group theory
analysis of this structure of elemental boron suggests that the tilt of the Dirac cone in its spectrum
can be tuned by a perpendicular electric field~\cite{Tohid2019Spacetime}

There has been many theory efforts to understand TDCMs. 
Kawarabayashi \etal~\cite{Kawarabayashi2011} find a generalized chiral symmetry that protects the Dirac node in these systems.
A generalized hopping model on honeycomb lattice supporting TDC is given by Kishigi \etal~\cite{Kishigi2011} as well 
as in quinoid-type graphene~\cite{Goerbig2008}. Initially motivated by the above organic compound, and later by the 2D 
$8Pmmn$ borophene, many physical properties of TDF systems are calculated. 
TDFs in 2D also exhibit the minimal conductivity phenomenon of upright Dirac fermions in graphene. In this case
$\sqrt{\sigma_{xx}\sigma_{yy}}$ approaches the same value as in graphene~\cite{Naumis2019MinimalCond}. 
Proskurin \etal study longitudinal conductivity of 2D TDCMs in magnetic fields and find
non trivial Landau levels transverse to tilt direction~\cite{Proskurin2015} and a divergent transverse conductivity~\cite{Proskurin2014}.
Rostamzadeh \etal using Boltzmann and Kubo formulas obtain the ratio of the transverse and longitudinal conductivities diverges as $\sqrt{1-\zeta^2}$~\cite{Rostamzadeh2019},
where $\zeta$ is the dimensionless quantity that determines the tilt. For type-I (II) TDFs one has $\zeta<1$ ($>1$). 
The effect of particle-hole asymmetry in the optical conductivity of TDFs was examined in~\cite{Tarun2017}.
Nishine \etal~\cite{Nishine2010} find cusps in dynamical polarization that leads to new plasmon modes~\cite{Nishine2011}.
The analytical results of Jalali-Mola \etal also show a kink in the plasmon dispersion,
and an additional over-damped plasmon mode arising from the tilt~\cite{Sahar2018a,Sahar2018b}.
The tilted is a very essential element in a minimal model that is able to generate a finite
quadruple moment ${\cal Q}_{ij}$~\cite{Gao2018Quadrapole}. 
It has been suggested that the tilt of the Dirac cone influences the spin transport~\cite{Sinha2019SpinTilted}.
Anomalous heat flow driven by the tilt was found by Sengupta \etal~\cite{Sengupta2018Anomalous} for 2D TDFs. 
Given the relevance of strong correlations in the organic TDF systems, the
effect of disorder and Coulomb interactions was also studied in 2D DFs~\cite{Isobe2012,Yang2018RGtilt}. 
The role of Coulomb interactions in generating dynamical (excitonic) gap in two dimensional TDCMs was investigated in~\cite{Xiao2017DynamicGap} where it was
found that the tilt suppresses the dynamic gap. Excitonic instability of the tilted Dirac fermions is also investigated by Ohki and coworkers~\cite{Ohki2019Excitonic}.
Tilt plays important role when superconductivity is introduced: Faraei \etal find that both retro and specular Andreev reflected holes come closer
to the normal to the interface upon increasing the tilt parameter and the Andreev reflection becomes perfectly perpendicular when $\zeta\to 1$~\cite{Faraei2019PAR}.
Furthermore in SNS junctions based on 2D TDFs, the Andreev mode that propagates along the channel will acquire an electric charge when the
tilt parameter is non-zero~\cite{Faraei2020Charged}. The pairing correlations are in general enhanced upon approaching the limit $\zeta=1$ that
separates the type-I and type-II Dirac/Weyl fermions~\cite{Rosenstein2017}. 

In three dimensions the tilted or type-II Weyl fermions in WTe$_2$ were originally predicted and noticed as 
their Weyl node was a protected as the meeting point of electron and hole pockets~\cite{Soluyanov2015}. This 
prediction was soon confirmed in experiment~\cite{Bruno2016WTe2,Wu2016WTe2,Wang2016WTe2}. Later on
MoTe$_2$ was also predicted~\cite{Sun2016MoTe2} and confirmed~\cite{Deng2016,Jiang2017MoTe2,Huang2016MoTe2,Xu2016MoTe2,Liang2016MoTe2} to be
a tilted Weyl material. The Fermi arcs in WTe$_2$ compound were observed in~\cite{Adam2016}. 
Later on, also type-II Dirac fermions were predicted in PtSe$_2$, PtTe$_2$, PdTe$_2$ and PtBi$_2$ family of 
dichalcogenides~\cite{Huang2016}. Single crystals of bulk PtSe$_2$ were grown and evidence for type-II Dirac cone was found in 
Ref.~\cite{Zhang2017PtSe2}. Fei \etal~\cite{Fei2017} found non-trivial Berry phase in layered type-II Dirac fermion in PdTe$_2$ crystals.
They also found that Pt alloying of IrTe$_2$ can shift the node to approach the type-II Dirac semimetal~\cite{Fei2018}.
The PdTe$_2$ compounds hosts both superconductivity and type-II Dirac fermions~\cite{Kim2017}. 
ARPES evidence for Lorentz violating type-II Dirac fermions was found in bulk PtTe$_2$~\cite{Yan2017}.
The inverse Perovskite compound Ca$_3$PbO also hosts 3D TDFs~\cite{Kariyado2011}. 
Subsulfide Ir$_2$In$_8$S is also shown to host type-II 3D Dirac fermions~\cite{Khoury2019}.

On the theory side, Trescher and coworkers found that a simple anisotropy affects the conductance, but not the Fano factor, while the tilt 
affects both Fano factor and conductance~\cite{Bergholtz2015}.
Effect of disorder was also found to be enhanced in three dimensional tilted Weyl materials~\cite{Bergholtz2017}. 
Landau quantization of 3D type-I and type-II Weyl semimetals was investigated by Tchoumakov \etal~\cite{Goerbig2016}.
Optical signature of the tilt in type-I and type-II Weyl semimetals was calculated by Carbotte~\cite{Carbotte2016Optical}.
In effect of disorder and Coulomb interactions was studied using renormalization group method in 3D TDFs~\cite{Lars2017RGtilt}.
It was found that the disorder enhances the effective tilt~\cite{Lars2017RGtilt}. 
Furthermore, the tilt was found to give rise to intrinsic anomalous Hall conductivity~\cite{Zyuzin2016}.\\

Finally, regarding possible materials for 1D TDFs, 
it has been suggested that the sodium termination of zigzag edges in graphene nano-ribbons can give rise
to 1D tilted Dirac cone spectrum~\cite{Ari2011Tilted1D}. 

\subsection{Tilted Dirac/Weyl materials as new spacetime structure in the solid state}
The list of materials exhibiting tilted Dirac/Weyl fermions
is being expanded in both theory and experimental fronts in $d=1,2,3$ dimensions. 
There are two approaches to such
systems. (i) The first approach is to take the tilt in the energy dispersion as granted. Then one can write an 
effective Hamiltonian compatible with the tilt and study its consequences. (ii) The second line of thought 
is to attribute the tilt in the dispersion relation to a new spacetime structure. This approach is pioneered
by Volovik in three dimensional Weyl semimetals and is followed by others. The later approach enjoys a 
covariant mathematical structure and necessitates the use of the geometric language of general relativity. 
Therefore, the powerful language of geometry can make  certain phenomena more manifest and/or transparent by expressing the
physics in a covariant and mathematically neat form. In this way, the plethora of phenomena associate with the structure of
spacetime can be examined in a solid-state setting. In this part of the introduction we would like to elaborate on this aspect
and review existing attempts in this direction:
As pointed out for Dirac/Weyl materials with upright cone,
the emergent structure of the spacetime felt by the electrons is the Minkowski spacetime. 
Tilting the cone-shaped dispersion spoils the Lorentz symmetry and hence the Minkowski structure of the 
emergent spacetime. Therefore a valid and pertinent question would be, what is the new spacetime structure
behind a tilted Dirac/Weyl cone spectrum?

This geometric line of thought in 3+1 dimensional Weyl semimetals is pioneered by Volovik~\cite{Volovik2016,Nissinen2017}. Indeed the
energy spectrum of Eq.~\eqref{TD.eqn} can be written as an invariant equation as $g^{\mu\nu}k_\mu k_\nu=m^2$,
where $m$ can be either zero or non-zero Dirac mass, $k_\mu=(E,\bsk)$ is the energy-momentum four-vector and
$g_{\mu\nu}$ is given by the so called Painelev\'e-Gullstrand (PG) metric~\cite{Tohid2019Spacetime},
\be
   ds^2=-v_F^2dt^2+(d\bs r-v_F\bs\zeta dt)^2 
   \label{PG.eqn}
\ee
In 3+1 dimensions this metric can be brought to the standard Schwarzschild format~\cite{Martel2001}
which admits a black-hole horizon~\cite{Carrol}. The explicit form of the above metric in 3+1 dimensions is
 \be
 g_{\mu \nu}=
 \begin{bmatrix}
  \zeta^2-1 & -\zeta_x & -\zeta_y & -\zeta_z\\
  -\zeta_x & 1 & 0 & 0 \\
  -\zeta_y & 0 & 1 & 0\\
  -\zeta_z & 0 & 0 & 1
 \end{bmatrix},
 \label{metric3.eqn}
 \ee
where $\zeta^2=|\bs\zeta|^2=\zeta_x^2+\zeta_y^2+\zeta_z^2$. 
In 2+1 dimensions, last column and row of the above matrix will be omitted. Note that embedding the two dimensional
graphene and deforming it by strain possible Lobachevsky space geometry can be constructed~\cite{IorioPRD2014,Iorio2013}. 
But the important difference of graphene metric with the 2+1 D version of metric~\eqref{metric3.eqn} is that
in graphene only the spatial components $g_{ij}$ can be influenced by strain, while in tilted Dirac/Weyl
materials the $g^{0j}$ components mixing space and time are subject to change. Furthermore, strain induced
changes are very typically very small effects. Formally the PG metric is basically a superposition of a Galilean 
boost on a Minkowski metric and can be realized in rotating frames which can lead to interesting coupling between
(quantum mechanical) spin and mechanical rotation~\cite{Maekawa2011PRB,Maekawa2011PRL}. 

Although the tilting deformation of the Dirac theory destroys the standard Lorentz symmetry,
but even for a uniform tilt parameter (i.e. a tilt parameter independent of spacetime coordinates), 
a deformed version of the Lorentz symmetry appears~\cite{Jafari2019}. Such a modified Lorentz symmetry
can be obtained as isometrics of the deformed Minkowski spacetime via standard mathematical procedure~\cite{Jafari2019}. 
Therefore, this symmetry can be attributed to a new spacetime structure. 
Indeed, the polarization function of tilted Dirac cone systems was shown to acquire a covariant form in the deformed 
Minkowski spacetime given precisely by metric~\eqref{metric3.eqn}~\cite{Sahar2019Polarization}.

Ojanen and coworkers propose that spatially varying time-reversal (TR) and inversion (I) breaking sources
in Weyl semimetals are equivalent to a curved spacetime for chiral fermions~\cite{Ojanen2019}. Such structures 
give rise to synthetic gauge fields. The present authors have proposed that in 2+1 dimensions, the 2D spatial
atomic arrangements allow to tune the geometry of the spacetime by electric fields~\cite{Tohid2019Spacetime}. 
Unlike strain induced changes in the metric of the spacetime, the changes introduced by TR or I breaking agents 
in (particularly 2D materials) is not a small effect. These ideas are further extended to meta-materials based on
Weyl semimetals by Ojanen and coworkers~\cite{OjanenPRX} in order to design the structure of the spacetime.  

In this paper we are interested in a 2D material hosting a tilted 2+1 dimensional tilted Dirac cone. 
There are two Dirac cones that in the inversion symmetric case are described by two tilted Dirac
cones with opposite tilt parameters $\bs\zeta$. Otherwise, their tilt parameters are arbitrary. 
In this paper we will study in detail the consequences of the spacetime dependent tilt parameter $\bs\zeta$ 
and will show the emergence of non-Abelian gauge fields that in the non-relativistic limit can be interpreted
as geometry induced spin-orbit couplings. We further obtain the effect of curvature on the semi-classical motion
(geodesics) and show that the tilt parameter gives rise to new forces which will be required in appropriate
extensions of the Boltzmann equation in such spacetimes.

\section{Non-Abelian gauge theory in tilted Dirac cone materials}
In this paper we will be interested in the Dirac materials with tilted conic spectrum in
two space dimensions. The metric of the resulting 2+1 D spacetime with tilt parameters $\bs\zeta=(\zeta_x,\zeta_y)$ is given by
\begin{eqnarray}
	&g_{\mu \nu} =\left[ \begin{matrix}
	\zeta^2-1 & -\zeta_x & -\zeta_y \\
	-\zeta_x & 1 & 0 \\
	-\zeta_y & 0 & 1 
	\end{matrix}
	\right],\nn\\
	&g^{\mu \nu} = [g_{\mu \nu}]^{-1} = \left[\begin{matrix}
	-1 & -\zeta_x & -\zeta_y \\
	-\zeta_x & 1-\zeta_x^2 & -\zeta_x \zeta_y \\
	-\zeta_y & -\zeta_x \zeta_y & 1-\zeta_y^2
	\end{matrix}
	\right]
	\label{metric.eqn}
\end{eqnarray}
where $\zeta^2 = \zeta_x^2 + \zeta_y^2$. The components $\zeta_x\equiv\zeta_1$ and $\zeta_y\equiv\zeta_2$ of the tilt
are assumed to have arbitrary functional dependence on the space coordinates $(x,y)$ inside the material. 
As a concrete example of how to generate such a space dependence in the tilt parameter $\bs\zeta$,
we have previously shown in $8pmmn$ borophene, that an external displacement field perpendicular to the 2D material couples to electronic degrees of
freedom in such a way that it controls the tilt parameters $\bs\zeta$~\cite{Tohid2019Spacetime}. Therefore a given space-dependent profile of perpendicular
electric field will imprint a corresponding profile of metric. This will in general amount to electric-field control of the
geometry of the spacetime. Another possible rout based on organic compounds would be the replacement of iodine with 
halogens~\cite{Geilhufe2018} in a space-dependent. 

Before studying the effect of a generic spacetime dependent entries $\bs\zeta$ in Eq.~\eqref{metric.eqn}
on the physical properties of TDCMs, let us recall the physics of strain in graphene. 
The strain in 2D materials can be formalized in terms of space-dependent metric entries. 
The resulting metric induced from a 3D Euclidean space to describe a deformed graphene gives rise to a 
curvature~\cite{Arias2015}. The Gaussian curvature of the deformed graphene will be equivalent to an effective (pseudo) 
magnetic field~\footnote{Here pseudo means that the sign of such magnetic field is opposite the two valleys.}. 
Therefore it is tempting to think that in the spacetime~\eqref{metric.eqn} of the TDCMs too, the role of spatial variation
in $\bs\zeta$ parameters will mimic an effective magnetic field. However, as we will show in this section,
allowing $\bs\zeta(x^\mu)$ to depend on spacetime coordinates, will generate 
non-Abelian gauge fields which in the non-relativistic limit correspond to various forms of spin-orbit 
interactions. This is unlike the Abelian (pseudo) gauge fields arising from strain in graphene. 

To understand the source of this difference, please note that in the case of strain fields in typical 2D materials,
the strain field affects the {\em spatial} components $g_{ij}$ of the resulting metric. Due to the Minkowski nature of the
parent graphene Hamiltonian, the strain field can not induce $g_{\mu 0}$ entries (off-diagonal entries mixing space and time). 
However, in the present case, the structure of the emergent metric~\eqref{metric.eqn} of the TDCMs is such that, allowing the $\bs\zeta$ to vary
in space, can only modify the off-diagonal components $g_{\mu0}$. The space part, $g_{ij}$ remains totally diagonal. 
As such, as long as there is no strain field to generate 
off-diagonal entries in the spatial part, $g_{ij}$ of Eq.~\eqref{metric.eqn}, we will not have any effective pseudo magnetic field 
corresponding to $U(1)$ gauge fields.
Then the question is, what type of forces are generated in this case?

To answer this question, we will need a brief reminder from standard geometry knowledge (see any standard textbook on general relativity, e.g. \cite{Ryder,Carrol,Fliessbach}). 
For a spacetime with arbitrary metric $g_{\mu\nu}$, the Christoffel symbols defined by~\cite{arfken},
\begin{eqnarray}
   \Gamma^{~~\rho}_{\mu \nu}= \frac{1}{2}g^{\rho \sigma}\left(\partial_\mu g_{\nu \sigma}+ \partial_\nu g_{\mu \sigma} - \partial_\sigma g_{\mu \nu}\right)
   \label{Gammadef.eqn}
\end{eqnarray}
are the essential entities that allow us to (i) construct "covariant" derivative, (ii) construct equations of motions of particles (geodesics), (iii)
to obtain the curvature tensor~\cite{Ryder,Carrol,Fliessbach}.  

\subsection{Emergent non-Abelian gauge fields}
As we pointed out, the first use of Christoffel symbols~\eqref{Gammadef.eqn} is to construct covariant derivatives. 
Assume that a vector is specified by its components $V^\mu$ (or $V_\mu$). The covariant derivative is given by,
\be
  \nabla_\nu V^\mu\equiv V^\mu_{;\nu}=V^\mu_{,\nu}+\Gamma^{\mu}_{\lambda\nu} V^\lambda,~~
  \nabla_\nu V_\mu\equiv V_{\mu;\nu}= V_{\mu,\nu}-\Gamma^{\lambda}_{\mu\nu} V_\lambda
  \label{nablamu.eqn}
\ee
where $V^\mu_{,\nu}\equiv \partial_\nu V^\mu$ is the partial derivative. Let us see where do we need to use these
derivatives. The Dirac equation in general is expressed by first oder derivative $(\gamma^\mu\partial_\mu+ m) \psi=0$.
In an arbitrary geometry, the curvature of spacetime requires to take the derivative of the spinor $\psi$ in a covariant form.
This amounts to replacement $\gamma^\mu\partial_\mu\to \gamma^a e^\mu_a (\partial_\mu+\Omega_\mu)$ where the $\Omega_\mu$ is called the "spin" connection,
as it needed to take covariant derivative of an "spinor"~\cite{Ryder,Carrol,nakahara,birrell}. 
When the parameters $\bs\zeta$ do not depend on space coordinate, the spin connection is $\Omega_\mu=0$ 
(for details see appendix~\ref{spin.app}). The frame fields $e^\mu_a$ will be required
to change the basis in a way that a locally flat Minkowski spacetime is obtained. 
When $\bs\zeta$ becomes space dependent, in addition to the basis change, one has to worry about the spin connection $\Omega_\mu$
which will be defined and computed below. The spin connection $\Omega_\mu$ is related to the connections defined in
Eq.~\eqref{nablamu.eqn} through the Christoffel symbols $\Gamma$'s as follows: For a generic manifold defined 
by metric $g_{\mu\nu}$ with $\mu,\nu=0,1,2$, there is a locally flat (Minkowski) manifold defined by metric $\eta_{ab}$ ($a,b=0,1,2$) 
tangent to this manifold. Requiring the spacetime length element in both cases to be identical gives,
\be
   g_{\mu \nu} = \eta_{ab}e_\mu^a e_\nu^b~~~\leftrightarrow~~~
   g^{\mu \nu} = \eta^{ab}e^\mu_a e^\nu_b,
   \label{vierbeindef.eqn}
\ee
which defines the frame fields $e_\mu^a$ (in 3+1 dimensions are called {\em vier}beins). From these frame fields one can then
construct, 
\be
   \omega_\mu ^{a b} = e^a_{\lambda}(x)g^{\lambda \sigma}(x)\nabla_\mu e^{b}_{\sigma}(x)
\ee
where the covariant derivative of the frame fields are  
\begin{eqnarray}
\nabla_\mu  e^a_{\sigma} = \partial_\mu e^a_{\sigma} -\Gamma_{\mu \sigma}^{~~\lambda} e^a_{\lambda}.
\end{eqnarray}
Finally employing the generators $\Sigma_{ab}=[\gamma_a,\gamma_b]/4$ of the Lorentz group, we can form the spin connection as
\be
   \Omega_\mu = \frac{1}{2}\omega_\mu ^{a b}\Sigma_{ab}.
   \label{Omega.eqn}
\ee

With the above quick reminder from geometry, we are now ready to compute the effect of arbitrary space dependence
of $\bs\zeta$ in 2+1 dimensional TDCMs. The first thing we need to do is to compute the Christoffel symbols $\Gamma$'s. 
It is straightforward, but cumbersome~\footnote{
There are plenty of well established algebraic manipulation programs to calculate the above symbols and much more
by just giving the functional form of the entries of the metric. For example see :~GRQUICK, https://library.wolfram.com} to use Eq.~\eqref{Gammadef.eqn} to explicitly obtain Eq.~\eqref{Gammas.eqn}. 
For details please see appendix~\ref{Gammas.app}.
Applying the steps outlined above to the metric~\eqref{metric.eqn} after a long but straightforward algebra 
we obtain the components $\omega_\mu^{ab}$ of the spin connection. The non-zero components will be given 
by Eq.~\eqref{spinconnexpanded.eqn} which can be compactly written as,
\bearr
   \omega_i^{~0j}=\alpha_i^{~j},~\omega_0^{~0i}=-\alpha_i^{~j}\zeta_j,~\mbox{others}=0
   \label{compactspinconnection.eqn}
\eearr
where the quantity $\alpha_i^{~j}\equiv (\partial_i\zeta^j+\partial^j\zeta_i)/2$
is suggested by Eq.~\eqref{spinconnexpanded.eqn} is the symmetric partial derivative of the tilt
$\bs\zeta$~\footnote{Note that $\zeta_i=\zeta^i$ are {\em parameters} of the spacetime metric, and the indices here are 
not raised or lowered by the metric itself. So the locations of indices in the right side of this equation does not matter 
and are just set to balance the location of the indices in the left side. This is manifest when comparing to Eq.~\eqref{spinconnexpanded.eqn}.}.

Equipped with these results, 
we are now ready to discuss the effect of the above spacetime structure in 2+1 dimensions. 
To proceed further, let us choose the following representation for the Clifford algebra, 
 \begin{eqnarray}
\gamma^0 = i\sigma_z,~~
\gamma^1 = \sigma_y,~~ 
\gamma^2 = -\sigma_x.
\label{Clifford1.eqn}
\end{eqnarray} 
Using the definition~\eqref{Omega.eqn} of the spin connection and performing the summation over $a,b=0,1,2$, only 
non-zero $\omega_\mu^{ab}$s contribute whereby we obtain,
\begin{eqnarray}
  &&\Omega_0 = \frac{1}{2}\omega_0 ^{a b}\Sigma_{ab}=-\frac{1}{4}\left[\alpha_1^j\zeta_j\sigma_x+\alpha_2^j\zeta_j\sigma_y \right]\nn\\
  &&\Omega_i = \frac{1}{2}\omega_i ^{a b}\Sigma_{ab} = \frac{1}{4} \left[ \alpha_i^1 \sigma_x  +\alpha_i^2 \sigma_y \right]
\end{eqnarray}
Once the spin connections $\Omega_\mu$ are computed, one can readily construct
the associated gauge fields~\cite{birrell,nakahara},
\begin{eqnarray}
\mathcal{A}_a = e^{~\mu}_a \Omega_\mu.
\end{eqnarray}
Using the explicit forms of the frame fields $e^{~\mu}_a$ given in appendix~\ref{spin.app}, we obtain
\begin{eqnarray}
&\mathcal{A}_0 = e^0_0 \Omega_0 + e^1_0 \Omega_1 + e^2_0 \Omega_2 = 0, \nonumber\\
&\mathcal{A}_1 = e^0_1 \Omega_0 + e^1_1 \Omega_1 + e^2_1 \Omega_2 = \Omega_1, \nonumber\\
&\mathcal{A}_2 = e^0_2 \Omega_0 + e^1_2 \Omega_1 + e^2_2 \Omega_2 = \Omega_2.
\label{gauges.eqn}
\end{eqnarray}
This equation establishes that the space dependence of the tilt parameters $\bs\zeta$ in metric~\eqref{metric.eqn},
induces non-Abelian gauge fields given by Eq.~\eqref{gauges.eqn}.

\section{Spin-orbit from curvature}
\label{so.sec}
In the context of graphene, it is well known that strain fields induce $U(1)$ gauge fields~\cite{vozmediano2010,voznediano2,guinea,guinea2}. 
As can be seen in Eq.~\eqref{gauges.eqn}, the components of the gauge field in the context of the spacetime structure~\eqref{metric.eqn} 
related to tilted Dirac materials have matrix structure which makes the non-Abelian gauge fields. In the context of 
solid state systems, the Pauli matrices $\sigma_i$ can denote the real spin (such as the helical states of a topological insulator)
of the pseudo-spins (as in the case of graphene). In this section we will show that the space dependence of $\bs\zeta$ will 
generate a coupling between the orbital motion and the (pseudo-)spin $\bs\sigma$. 
To develop an intuition for the meaning of such gauge fields, it is useful to consider a massive Dirac particle with $m\ne 0$
that allows to study the "non-relativistic limit" of the underlying Dirac system.
Such a mass term does not change the structure of the metric. The only modification arising
from the mass term will appear in the right side of the dispersion relation $g_{\mu\nu}k^\mu k^\nu=m^2$. 

The meaning of the gauge structure~\eqref{gauges.eqn} can be best understood by expanding the Dirac hyperbolas around the band minima
and approximating them by parabolas. Therefore in tilted Dirac/Weyl systems too, the states near the
bottom of the conduction band (or top of the valence band) of the Dirac dispersion can be approximated by a parabolic band structure. 
In the context of the standard model of particle physics, this corresponds to non-relativistic limit where velocities are much less than
the upper limit of velocities (in our case the Fermi velocity, $v_F$). In this limit a Dirac-Foldy-Wouthuysen transformation
reveals how the spin-orbit interaction emerges from the Dirac equation. In our case, the same procedure will lead to 
a rich structure of pseudospin-orbit coupling. 

The non-relativistic limit of the tilted Dirac equation will be given by,
\begin{eqnarray}
\frac{1}{2m}\left(\bs p\bm 1 - g \bs{\mathcal A} \right)^2
\end{eqnarray}
where $\bs{\mathcal A}$ is given by Eq.~\eqref{gauges.eqn}. 
Expanding the above expression, one generates three type of terms, (i) $\bs{\mathcal A}.\bs{\mathcal A}$, 
(ii) $\bs{\mathcal A}.\bs p$ and (iii) $\bs\partial.\bs{\mathcal A}$ terms. They are given by the following
expressions:
\begin{eqnarray}
  \bs{\mathcal A}.\bs{\mathcal A} =& \frac{1}{16}\left[(\alpha_1^1 + \alpha_1^2)^2 + (\alpha_2^2 + \alpha_1^2)^2\right]=\sum_{i,j,k}\alpha_i^j\alpha_i^k \nn\\
  2\bs{\mathcal A}.\bs p =&\frac{1}{4}\left[ \alpha_1^1 p_x \sigma_x + \alpha_2^2p_y \sigma_y + \alpha_1^2 (p_x\sigma_y+p_y\sigma_x) \right]\nn\\
 -i\bs \partial \cdot  \bs{\mathcal A} =& \frac{1}{4}\left[(\partial_i\alpha_1^i)\sigma_x+(\partial_i\alpha_2^i)\sigma_y \right]
\end{eqnarray}
The non-relativistic limit of this Hamiltonian becomes,
\be
   \frac{1}{2m}\left[\bs p^2 +g\bs{\cal A}\cdot\bs{\mathcal A}+g \bs d.\bs\sigma+g \bs b.\bs\sigma \right]
\ee
where,
\bearr
   d_{n}(\bs p)=\alpha_n^j p_j,~~~~b_n=-i\partial_j\alpha^j_n,
   \label{d.eqn}
\eearr
where $n,j$ run over the spatial indices $1,2$. 
This analysis clearly shows that the non-Abelian gauge potential, in the non-relativistic limit
corresponds to two contributions: (i) The $\bs d$ term that directly couples momentum and (pseudo) spin, 
is the emergent spin-orbit coupling that arises from the coordinate dependence of the $\bs\zeta$ that parametrizes
the metric~\eqref{metric.eqn}. (ii) The $\bs b$ term  $g\bs b.\bs\sigma$ is emergent Zeeman term. Note that,
this must be distinguished from the curvature induced $U(1)$ field that couples to 
to orbital motion of the electrons can generate Landau quantization. This term being a Zeeman-like term, can only couple to the
(pseudo-) spin degree of freedom. Note that both $\bs d$ and $\bs b$ terms are odd functions of $\bs\zeta$. 
Therefore, in an inversion symmetric Dirac material where the one can toggle between the two valleys by
$\bs\zeta\to -\bs\zeta$, this Zeeman-like term also changes sign and are therefore {\it pseudo}-Zeeman like terms.

Mathematically, the field $\zeta^i\equiv \zeta_i$ being a vector field either has a zero circulation,
or non-zero circulation. When its circulation is zero, it can be written as,
\begin{eqnarray}
   \zeta_i= \partial_i \Phi 
   \label{Helmholtz.eqn}
\end{eqnarray}
In this case, the important parameters $\alpha_n^j$ are given by, $\alpha_n^j=\partial_n\partial_j\Phi$
which then for $d_{n}$ and $b_n$ gives,
\bearr
 d_{n}(\bs p) =  \partial_n  \bs p \cdot \bs \partial \Phi,~~~~b_n=-i\partial_n \partial^2 \Phi
 \label{dn.eqn}
\eearr
For an inversion symmetric material, if one valley has a tilt parameter $\bs\zeta$, the other valley must have $-\bs\zeta$. 
Therefore it is reasonable to assume that in real space too, the field $\bs\zeta$ arises from sinks and
sources of equal charges $\pm Q$~\footnote{Note that here $Q$ is a topological charge associated with the
tilt field $\bs\zeta$.}. Therefore, except for isolated pairs of points, the condition $\bs\partial\cdot\bs\zeta=0$ is satisfied.
The implication of this condition on $\Phi$ is,
$\partial^2\Phi=\pm Q\delta(\bs r-\bs r_\pm)$. Choosing a linear combination of Harmonic functions to satisfy the 
boundary condition $\bs\partial \Phi(\bs r\to\infty)=\bs\zeta_\infty$ (assuming a flat spacetime profile 
specified by $\bs\zeta_\infty$) gives $\Phi=\bs\zeta_\infty.\bs r+Q\ln(|\bs r-\bs r_\pm|)$. 
The charges $\pm Q$ are integer topological charges of the field $\bs\zeta$. Non-zero values of $Q$ when 
inserted in Eq.~\eqref{dn.eqn} give a singular $\partial_n Q\delta(\bs r-\bs r_\pm)$.
Therefore the topological index $Q$ governing $b_n$, can not be non-zero. The only remaining term $\bs \zeta_\infty.\bs r$ when inserted
in Eq.~\eqref{dn.eqn} will give zero. The conclusion is that, if $\bs\partial\times\bs\zeta=0$, both the spin-orbit coupling
and the curvature induced Zeeman fields vanish. 

Now let us consider the second possibility, namely a non-zero $\bs\partial\times \bs\zeta$. A general enough 
choice of $\bs\zeta=\rho f(\rho) \hat\varphi$ where $\hat\varphi$ is the unit vector in cylindrical coordinate
corresponding to azimuthal angle $\varphi$, and $\rho$ is the distance from the origin. This choice gives
\be
   \bs\partial\times\bs\zeta=\frac{1}{\rho}\frac{\partial}{\partial\rho}\left[\rho^2 f\right] \hat z
   \label{curlzeta.eqn}
\ee
In this case the important parameters $\alpha^j_n$ are given by,
\be
   \rho^{-1}f' \left(x_n\eps_{j\ell}x_\ell+x_j\eps_{n\ell} x_\ell \right)
\ee
where $\eps_{12}=-\eps_{21}=1$ and $\eps_{11}=\eps_{22}=0$ defines the Levi-Civita symbol and $f'=\partial_\rho f$. 
Inserting the above result in Eq.~\eqref{d.eqn} gives,
\bearr
   d_1&=f'\rho(p_x\sin2\varphi+p_y\cos2\varphi)/2\nn\\
   d_2&=f'\rho(p_x\cos2\varphi-p_y\sin2\varphi)/2\label{df.eqn}\\
   b_1&=(-i) \rho\sin\varphi\left(\rho g'\sin^2\varphi-3g\right)\nn\\
   b_2&=(-i) \rho\cos\varphi\left(-\rho g'\cos^2\varphi-3g\right)\label{bf.eqn}
\eearr
where we define $g=f'/\rho$. Adopting a representation for the Clifford algebra that differs by the one in Eq.~\eqref{Clifford1.eqn}
in $\sigma_x\leftrightarrow \sigma_y$, for the spin-orbit coupling term we obtain
\be
   f'\rho\left(\cos2\varphi \bs\sigma\cdot\bs p-\sin2\varphi \bs\sigma\times\bs p \right)/2
\ee
An important feature of the above spin-orbit coupling is its highly anisotropic nature. The nice-looking terms
$\bs\sigma\cdot\bs p$ is actually the Dresselhaus spin-orbit coupling. To see this one can transform back to the old
representation~\eqref{Clifford1.eqn}. The second term, $\bs\sigma\times\bs p$ in the new representation is clearly seen 
to be a Rashba spin-orbit coupling. Therefore, the space dependence of the tilt parameter $\bs\zeta$, whose circulation
is non-zero (determined by $f$) gives rise to Dresselhaus and Rashba spin-orbit couplings that are,
(i) highly anisotropic, (ii) their existence depends on the value of $f'$. Therefore a $\rho$-independent $f$, despite
giving rise to a non-zero constant circulation in Eq.~\eqref{curlzeta.eqn} will have $f'=0$ and hence both $\bs d$ and $\bs b$ vanish. 
This feature distinguishes the pseudo-Zeeman and pseudo-spin-orbit coupling that are generated by space dependence of the
tilt parameter $\bs\zeta$ from the other forms of spin-orbit coupling arising from rotation~\cite{Maekawa2011PRB,Maekawa2011PRL,Shitade2020}. 

\begin{figure}
	\centering
	\includegraphics[width =0.99 \linewidth]{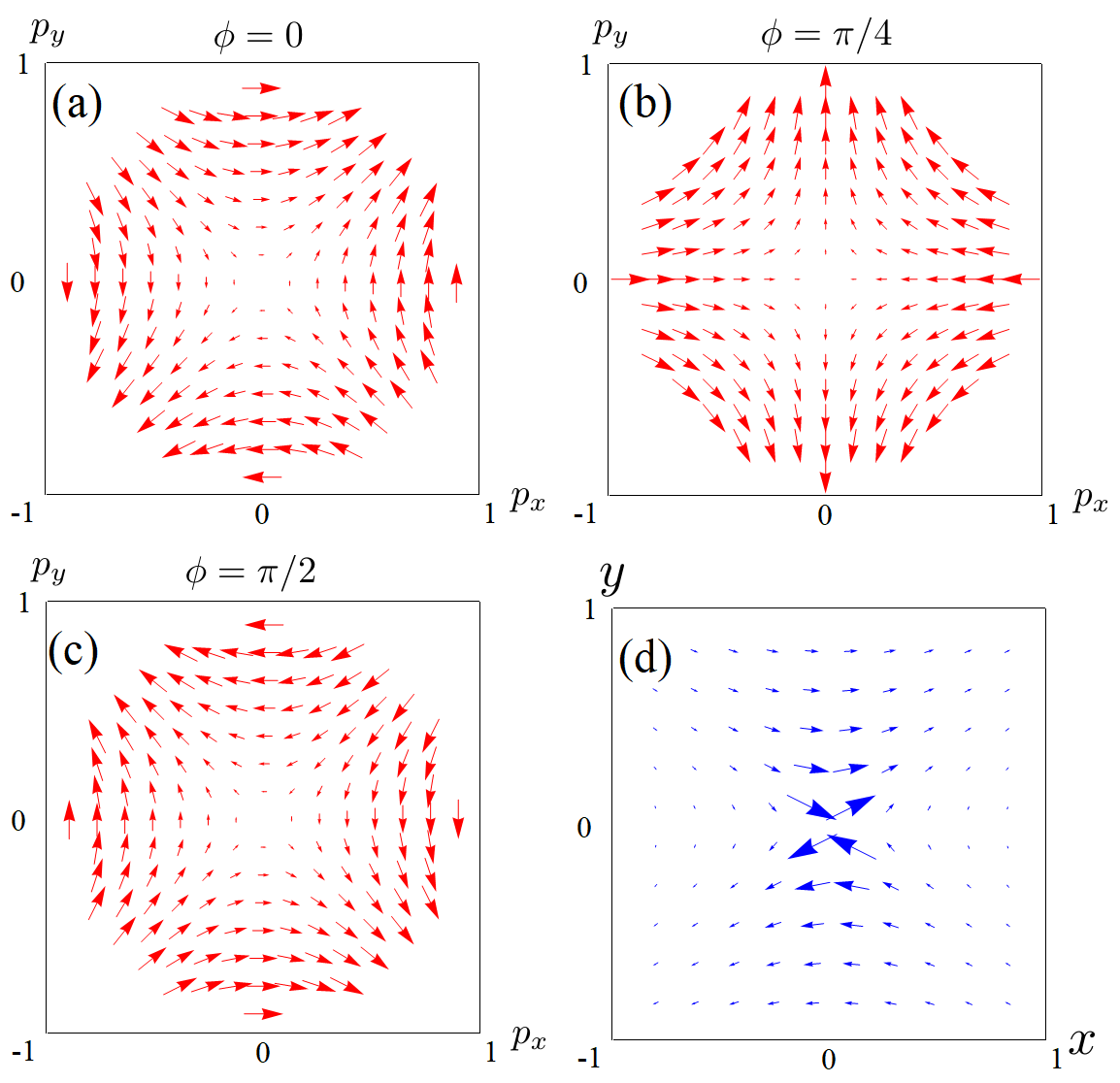}
	\caption{(Color online) Vector plot of spin-orbit coupling and pseudo-Zeeman field. (a-c) By choosing $f(\rho)=\ln\rho$ which is equivalent to $\bs\zeta=2\rho\ln\rho\hat\varphi$, girves rise to $f'\rho=2$ which eliminates the radial dependence of the spin-orbit coupling $\bs d$, plotted when $\phi = 0, \pi/4 , \pi/2$.
		 (d) Choosing $g=2\rho^{-2}$ leave a $1/\rho$ radial dependence in the pseudo-Zeeman field $\bs b$.}
	\label{plot}
\end{figure}

As an example, choosing $f(\rho)=\ln\rho$ which is equivalent to $\bs\zeta=2\rho\ln\rho\hat\varphi$, gives rise
to $f'\rho=2$ which eliminates the radial dependence of the spin-orbit coupling $\bs d$ in Eq.~\eqref{df.eqn}.
Correspondingly we have $g=2\rho^{-2}$ which will leave a $1/\rho$ radial dependence in the pseudo-Zeeman field $\bs b$.
These features are plotted in Fig.~\ref{plot}. 
The above analysis in the small $\bs p$ reveals the structure of spin-orbit coupling (or pseudospin-orbit coupling if $\bs\sigma$
is not the real spin) arising from space-dependent tilt parameter $\bs\zeta$. When the $\bs\zeta$ that enters as an off-diagonal
term mixing space and time is absent, as can be seen from Eq.~\eqref{compactspinconnection.eqn}, 
the $\omega^{i0}_\mu$ components of the spin-connection vanish from which it follows that the non-Abelian gauge fields
in Eq.~\eqref{gauges.eqn} vanish. When the $\bs\zeta$ term is introduced in tilted Dirac/Weyl systems and for whatever reason possess 
a non-trivial space dependence, the connections $\omega^{0i}_\mu$ (which multiply the matrices $\gamma_0\gamma_i$) become
non-zero and generate the non-Abelian gauge structure in Eq.~\eqref{gauges.eqn}. 
Allowing the spatial components of the metric to depend on space (driven e.g. by strain), will generate
the $\omega^{ij}_\mu$ components which are well known in the context of graphene and can only generate the
$U(1)$ gauge structure, as they couple to $\gamma_i\gamma_j$ combination of Dirac matrices \cite{vozmediano2010,voznediano2}.

\section{Classical Geodesics}
When the length scale of applied fields are much larger than the spread of the wave packets, and the wave packets
themselves are larger than the lattice constant, the semi-classical Boltzmann transport can be used to study the
electron dynamics. In tilted Dirac/Weyl materials too, such a regime does exist and it is therefore appropriate to
study the classical geodesics in a generic background metric~\eqref{metric.eqn} with space dependent $\bs\zeta$. 
This will allow us to develop a feeling at the classical level to the nature of the forces that correspond to the the non-Abelian gauge structure~\eqref{gauges.eqn}.

It is a well known textbook fact that gravitational forces around a rotating source that only slightly deform the metric $\eta^{\mu\nu}$ of the
flat spacetime, can be effectively described by forces that resemble the electromagnetic forces. These effects go under
the name of gravitomagnetic effects~\footnote{See e.g. chapter 6 of the Ryder's textbook~\cite{Ryder}.}. 
The effect of rotation goes into a weak off-diagonal element that mixes space and time. This feature is 
similar to our metric~\eqref{metric.eqn}, except that (i) $\bs\zeta$ in our case is  not necessarily weak and (ii) 
its functional dependence on space coordinates $(x,y)$ can in principle be anything. 
Therefore in this section we set out to understand the meaning of the space-dependent tilt parameter $\bs\zeta$ in the metric~\eqref{metric.eqn}
of the tilted Dirac/Weyl materials. We will follow the textbook approach of Ryder and will start by writing down 
the geodesics equations in the background metric~\eqref{metric.eqn} in order to identify
the structure of the new forces that arise from spatial variation of $\bs\zeta$ at the {\it classical level}.
In the case of graphene where only the spatial components $g^{ij}$ of the metric are allowed to depend on space (due to strain field),
it has been found that the curvature induced $U(1)$ forces, lead to spatial separation of the valley currents~\cite{Szpak}. 

Our classical treatment in this section, parallels the quantum treatment of section~\ref{so.sec} as in both 
cases we consider the limit where a mass term is present that allows to approximate the Dirac hyperbola by 
a parabola. The geodesic equation is given by~\cite{Ryder}
\begin{eqnarray}
\frac{d^2x^\mu}{d s^2} + \Gamma^\mu_{~~\nu \lambda} \frac{d x^\nu}{d s} \frac{d x^\lambda}{d s}=0
\end{eqnarray}
where $s$ is the proper time. In the non-relativistic (or in gravitational jargon the "Newtonian") limit, $s\approx t$
and $dx^0/ds\approx v_F$ and $dx^i/ds \approx v\ll v_F$.  In this limit the acceleration $a^i=d^2x^i/dt^2$ will become, 
\begin{eqnarray}
a^i & =-\left(v_F^2\Gamma^i_{~00} +2v_F \Gamma^i_{~0 k} v^k+ \Gamma^i_{~km} v^kv^m \right) \nonumber\\
&+\left(v_F^2\Gamma^0_{~00}  +2v_F \Gamma^0_{~0 k} v^k + \Gamma^0_{~km} v^kv^m \right)\frac{v^i}{v_F}.
\label{novterms.eqn}
\end{eqnarray}
Performing the summations over $k,m$ the separate components $i=1,2$ of the acceleration can be organized in 
powers of $v/v_F$ where $v$ is the band velocity near the bottom of conduction band and $v_F$ is the asymptotic
velocity of Dirac electrons at large momenta. The leading order is 
\begin{eqnarray}
\bs a_{(0)} = \frac{v^2_F}{2}\left(\bs\partial\zeta^2-\bs\zeta \bs\zeta \cdot \bs\partial \zeta^2 \right)  
\end{eqnarray}

The first term in this equation is a gradient term. If $\zeta^2$ can be interpreted as negative of an 
"electrostatic" potential, then the first term will correspond to the electric field arising from such term. 
There is however additional structure in the second term. When the variations of $\bs\zeta$ are purely transverse,
the longitudinal derivative $\bs\zeta\cdot\bs\partial$ will vanish. If the spatial profile of $\zeta^2$ generated by 
external fields is localized around some origin~\footnote{Perhaps this can be achieved by applying the
electric field via a tip of thickness of several tens of nanometers.}, and decays away from it, will correspond to an inward acceleration if
$m$ is positive (i.e. we for states near the bottom of conduction band). The gate voltage can be used to tune the 
chemical potential and therefore enables us to tune the Fermi level between the conduction and valence bands. This process will
change the sign of $m$. Therefore if the above term is attractive for electrons it will be repulsive for holes,
and vice versa. 

The first order contribution is,
\begin{eqnarray}
 a^i_{(1)}&=&v_F(\vec\zeta \cdot \vec \partial \zeta^2 -  \partial_{\bar i}\zeta_{\bar i}^3- \frac{1}{3} \partial_i\zeta_i^3 )v^i \nn\\
&&+ v_F v^{\bar i} \zeta_i(\vec \zeta \cdot \vec \partial \zeta_{\bar i} + \frac{1}{2}\partial_{\bar i}\zeta^2)\nn\\
&&+ v_F \left(\bs v\times \partial \times \bs \zeta\right)^i
\label{vterms.eqn}
\end{eqnarray}
where $i=1,2$ correspond to $\bar i=2,1$, respectively. 
Note that $i$ is the free index of the left side and there is no sum over $i$ on the right side. 

For the inversion symmetric tilted Dirac materials the two valleys have opposite $\bs\zeta$. 
The velocity independent terms, $\bs a_{(0)}$ in Eq.~\eqref{novterms.eqn} are even in $\bs\zeta$. Therefore, they 
do not change upon $\bs\zeta \to -\bs\zeta$. Therefore these terms are the same in both valleys in such materials. 
The first order terms in Eq.~\eqref{vterms.eqn} are odd in $\bs\zeta$. Therefore these can be attributed to the pseudo-forces in 
inversion symmetric system. Particularly,  
the third line in Eq.~\eqref{vterms.eqn} is a Lorentz-type force provided $\bs\zeta$ can be imagined as the
spatial part of a vector potential such that $\bs\partial\times \bs\zeta$ would correspond to a pseudo-magnetic field. Again by pseudo
we mean that this field begin odd in $\bs\zeta$, changes sign in the other valley. 
The first line resembles a "friction" force as the $i$'th component of the acceleration is proportional to $v^i$. 
Depending on the sign of the term in the parenthesis, this can be friction or anti-friction! In the later case, 
this term will cause an increase in the velocity, until the Newtonian regime ceases to be valid. In analogy with the
first line, one can think of the second line as a kind of transverse friction" force which is then expected to 
enhance the shear viscosity when the interactions are turned on to form an electron liquid. 

To develop a feeling for the classical dynamics in the spacetime~\eqref{metric.eqn}, let us 
consider the special case where $\bs \zeta = (h(x),0)$ is varied only unidirectionally along the $\bs\zeta$ direction.
We will further assume that $h(x)=\zeta_0 \tanh(\frac{x}{\lambda})$. 
In this case for the zeroth order contribution we get,
\begin{eqnarray}
a^1_{(0)} &&= \frac{v_F^2}{2}\left(\partial_x \zeta^2 - \zeta_x(\zeta_x\partial_x+\zeta_y\partial_y)\zeta^2 \right)\nn\\
&&= \frac{v_F^2}{2} h(x) h'(x) \left(1-h^2(x)\right) \nn\\
&&= \frac{v_F^2}{2} \frac{\zeta_0^3}{\lambda} \tanh (\frac{x}{\lambda}) \mbox{sech}^2(\frac{x}{\lambda})(1-\zeta_0^2 \tanh^2(\frac{x}{\lambda}))
\end{eqnarray}
and 
\begin{eqnarray}
a^2_{(0)} = \frac{v_F^2}{2}\left(\partial_y \zeta^2 - \zeta_y(\zeta_x\partial_x+\zeta_y\partial_y)\zeta^2 \right)=0
\end{eqnarray}
and the first order contribution is
\begin{eqnarray}
a^1_{(1)} = v_Fh'(x)h^2(x) v_1,~~~~a^2_{(1)} =- v_Fh'(x)h^2(x) v_2
\end{eqnarray}
which for our step-like function becomes,
\begin{eqnarray}
(a^1_{(1)},a^2_{(1)}) = v_F\frac{\zeta_0^3}{\lambda}\mbox{sech}^2(\frac{x}{\lambda}) \tanh^2(\frac{x}{\lambda}) (v^1,-v^2).
\end{eqnarray}

Note that in this section what we have calculated is the acceleration. To convert it to 
the force, one must note that it has to be multiplied by a mass term $m$. This term is
positive for the states near the bottom of the conduction band. For those near the
top of the valence band, this term (determining the parabola) is negative. 
Therefore the sign of the above forces can be reversed by e.g. gate
doping and toggling the chemical potential between the valence to conduction bands. 
In this way, the sign of the force depends on the sign of charge carriers. 
Therefore the electric and magnetic forces emerging from the structure of the spacetime~\eqref{metric.eqn}
reverse their signs by changing the "charge" of the carriers. As such, these forces can be
regarded as a natural generalizatioins of electric and magnetic forces that arise from
the geometry of the spacetime itself. The {\it external} electric or magnetic fields 
are assumed to be absent here. Therefore in studying the effect of external electric and magnetic
fields on the transport of electrons, one must in addition to external electric and magnetic fields
worry about the forces that arise from the curved nature of spacetime~\eqref{metric.eqn}. 

For a fixed chemical potential corresponding to a given sign of the energy, 
the sign of $m$ is fixed. In that case, all the terms, including the term that resebles the Lorentz-force 
in the third line of Eq.~\eqref{vterms.eqn} change sign. As such the classical Landau orbits for
electrons in the two valleys have opposite directions. Therefore this term (when strong enough)
can generate valley-polarized edge currents. So the curvature engineering can in principle
generate valley Hall effect. Even if the material is not perfectly inversion symmetric, the
symmetric part of it (for whith two valleys are related by $\bs\zeta\to -\bs\zeta$) is capable of generating valley polarized effects.

\section{Discussions and summary}
In this work we have investigated the effect of spatial dependence in the tilt parameter $\bs\zeta$ that determines the
metric~\eqref{metric.eqn}. The essential quantity is $\alpha_i^j$ defined under Eq.~\eqref{compactspinconnection.eqn}. 
At quantum level this quantity gives rise to non-Abelian gauge fields. The meaning of such gauge fields becomes clear 
in the non-relativistic limit pertinent the bottom of conduction or top of valence band states which corresponds to 
various forms of spin-orbit coupling. This agrees with recent proposal by Shitade and coworkers on geometric spin-orbit coupling ~\cite{Shitade2020}.
At the classical level, from geodesic equations one can infer various forms of
forces that have no analogs in solid state systems with Galilean structure. The sign of these forces can be changed by 
toggling between conduction and valence bands via a gate voltage and therefore for opposite charge carriers they have
opposite signs. This means that these forces are natural extensions of electric and magnetic forces to the geometry~\eqref{metric.eqn}.
In studying transport properties of such systems, these forces are also expected to play role in addition to the
{\em external} electric and magnetic fields acting on the charge carriers. 

It is worth to reiterate why unlike the strain induced pseudo-gauge forces in graphene, here we have a non-Abelian 
gauge structure. The reason is that strain appears in spatial components $g_{ij}$ while the space dependence of the tilt 
in TDCMs appears through the off-diagonal entries $g_{0\mu}$ that mix space and time coordinates. 
These two ways of modifying the $g_{\mu\nu}$ metric have two different physics. Therefore strain can be used as additional control parameter
to generate a pseudo-magnetic field in addition to the spin-orbit interactions arising from the spatial variations
of $g_{0\mu}$ entries of the metric in TDCMs.
Indeed generation of spin-orbit couplings by spatial variation of $\bs\zeta$ seems quite plausible: In our
previous work~\cite{Tohid2019Spacetime} we have found that the displacement field couples to $\bs\zeta$ via
Rashba spin-orbit coupling. Now here we have a reversed situation. If for whatever reason the tilt $\bs\zeta$ acquires 
a dependence on space coordinates $x^i$, one will have various spin-orbit couplings, including Rashba. 

In terms of possible materials realizations, in 2D materials there are many possibilities. 
The most prominent example is the organic compound. Replacing the iodine (I, $\zeta<1$) by halogens, such as
F ($\zeta>1$) an average control of F substitution correspond to a space dependent tilt~\cite{Geilhufe2018}.
Even a random substitution of halogens can lead to randomness in the $\bs\zeta$ which seems to be an interesting 
element to randomize and worth investigation, particularly from the point of view of the spacetime structure. 
Or the interface between I-rich and F-rich compound based on this organic systems can mimic black-hole horizon.
The transition between type-I and type-II Weyl fermions is signaled by an enhancement of the superconducting pairing~\cite{Rosenstein2017}.
This transition corresponds to crossing a black-hole horizon~\footnote{
One has to be careful that the "effective" description in terms of a metric is valid at length scales much larger than the
atomic scales. As such, such solid state black-holes are escapable in the atomic scales~\cite{Bergholtz2020}. }
In $8Pmmn$ borophene structure a displacement field couples to $\bs\zeta$ and can in principle be used to
imprint varous profiles of $\zeta_i(x,y)$~\cite{Tohid2019Spacetime}. Furthermore, magnetic textures~\cite{Ojanen2019} can also 
generate space dependet tilt parameter $\bs\zeta(x^i)$. 

Therefore the tilted Dirac cone systems in two space dimensions are promising frameworks for generation of spin-orbit coupling (synthetic gauge
fields). The effect of spacetime geometry is not limited to such synthetic forces. The quantum emission which is at the heart of solid-state spectroscopies 
will also be affected by the spacetime curvature~\cite{FKonig}. This might require a careful examination of the linear response theory that links
theoretical calculations within the Kubo formula with experiments~\cite{GirvinBook}.

\section{acknowledgments}
We wish to thank Ahmad Reza Moradpour and Armin Ghazi for fruitful discussions and Prof. Dr. Reza Mansouri for insightful
discussions and encouragements to push the idea of solid-state spacetime structures. S.A.J. appreciates
research deputy of Sharif University of Technology, Grant No. G960214 and Iran Science Elites Foundation (ISEF). 

\appendix

\section{Derivation of Christoffel symbols}
\label{Gammas.app}
In this appendix we present details of calculations related to the Christoffel symbols defined by, 
\begin{eqnarray}
   \Gamma^{~~\rho}_{\mu \nu}= \frac{1}{2}g^{\rho \sigma}\left(\partial_\mu g_{\nu \sigma}+ \partial_\nu g_{\mu \sigma} - \partial_\sigma g_{\mu \nu}\right),
\end{eqnarray} 
where the metric of the resulting 2+1 D spacetime with tilt parameters $\bs\zeta=(\zeta_x,\zeta_y)$ is given by
\begin{eqnarray}
&g_{\mu \nu} =\left[ \begin{matrix}
\zeta^2-1 & -\zeta_x & -\zeta_y \\
-\zeta_x & 1 & 0 \\
-\zeta_y & 0 & 1 
\end{matrix}
\right],\\
&g^{\mu \nu} = [g_{\mu \nu}]^{-1} = \left[\begin{matrix}
-1 & -\zeta_x & -\zeta_y \\
-\zeta_x & 1-\zeta_x^2 & -\zeta_x \zeta_y \\
-\zeta_y & -\zeta_x \zeta_y & 1-\zeta_y^2
\end{matrix}
\right]
\end{eqnarray}
where $\zeta^2 = \zeta_x^2 + \zeta_y^2$. The components $\zeta_x\equiv\zeta_1$ and $\zeta_y\equiv\zeta_2$ of the tilt
are assumed to have arbitrary functional dependence on the space coordinates $(x,y)$ inside the material. Then Christoffel symbols are given by
\begin{widetext}
\begin{align}
&\Gamma^{0}_{00} = \zeta_x \zeta_y (\partial_y\zeta_x+\partial_x\zeta_y)+\zeta_x^2\partial_x\zeta_x + \zeta_y^2\partial_y\zeta_y \nonumber\\
&\Gamma^{0}_{10}=\Gamma^{0}_{01} = -\zeta_x\partial_x\zeta_x-\frac{1}{2}\zeta_y (\partial_y\zeta_x+\partial_x\zeta_y) \nonumber\\
&\Gamma^{0}_{20}=\Gamma^{0}_{20} =-\zeta_y\partial_y\zeta_y -\frac{1}{2}\zeta_x (\partial_y\zeta_x+\partial_x\zeta_y) \nonumber\\
&\Gamma^{0}_{12}=\Gamma^{0}_{21} = \frac{1}{2} (\partial_y\zeta_x+\partial_x\zeta_y) \nonumber\\
&\Gamma^{0}_{11} = \partial_x\zeta_x \nonumber\\
& \Gamma^{0}_{22} = \partial_y\zeta_y \nonumber\\
&\Gamma^{1}_{00} = \zeta_x\zeta_y (\zeta_x\partial_y\zeta_x+\zeta_y\partial_y\zeta_y)+(\zeta_x^2-1)(\zeta_x\partial_x\zeta_x+\zeta_y\partial_x\zeta_y)\nonumber\\
&\Gamma^{1}_{01} =\Gamma^{1}_{10} = -\frac{1}{2}\zeta_x(2\zeta_x\partial_x\zeta_x+\zeta_y(\partial_y\zeta_x+\partial_x\zeta_y)) \nonumber\\
&\Gamma^{1}_{02} =\frac{1}{2} ( (1-\zeta_x^2)\partial_x\zeta_y-  (\zeta_x^2+1)\partial_y\zeta_x)-\zeta_x\zeta_y\partial_y\zeta_y \nonumber\\
&\Gamma^{1}_{21}=\Gamma^{1}_{12} = \frac{1}{2}\zeta_x (\partial_y\zeta_x+\partial_x\zeta_y) \nonumber\\
&\Gamma^{1}_{22}=\zeta_x\partial_y \zeta_y \nonumber\\
&\Gamma^{1}_{11} = \zeta_x\partial_x \zeta_x  \nonumber\\
&\Gamma^{2}_{00} =(\zeta_y^2-1)(\zeta_x\partial_y\zeta_x + \zeta_y\partial_y\zeta_y) + \zeta_x\zeta_y (\zeta_x\partial_x\zeta_x+\zeta_y\partial_x\zeta_y) \nonumber\\
& \Gamma^{2}_{01}=\Gamma^{2}_{10} = -\frac{1}{2} (\zeta_y^2-1)\partial_y\zeta_x-\frac{1}{2}(\zeta_y^2+1)\partial_x\zeta_y - \zeta_x\zeta_y\partial_x\zeta_x \nonumber\\
& \Gamma^{2}_{02} = \Gamma^{2}_{20} =-\frac{1}{2}\zeta_x \zeta_y (\partial_y\zeta_x+\partial_x\zeta_y)-\zeta_y^2\partial_y\zeta_y  \nonumber\\
& \Gamma^{2}_{21}=\Gamma^{2}_{12} =\frac{1}{2} \zeta_y (\partial_y\zeta_x+\partial_x\zeta_y) \nonumber\\
& \Gamma^{2}_{22}=\zeta_y\partial_y\zeta_y \nonumber\\
&\Gamma^{2}_{11} = \zeta_y\partial_x\zeta_x
\label{Gammas.eqn}
\end{align}
 \end{widetext}

\section{Derivation of the spin connection}
\label{spin.app}
In an arbitrary geometry, the curvature of spacetime requires to take the derivative of the spinor $\psi$ in a covariant form.
This amounts to replacement $\gamma^\mu\partial_\mu\to \gamma^a e^\mu_a (\partial_\mu+\Omega_\mu)$ where the $\Omega_\mu$ is called the "spin" connection,
as it needed to take covariant derivative of an "spinor"~\cite{Ryder,Carrol,nakahara,birrell}. The frame fields $e^\mu_a$ will be required
to change the basis in a way that a locally flat spacetime is obtained. For a generic manifold defined 
by metric $g_{\mu\nu}$ with $\mu,\nu=0,1,2$, there is a locally flat (Minkowski) manifold defined by metric $\eta_{ab}$ ($a,b=0,1,2$) 
tangent to this manifold. Requiring the spacetime length element in both cases to be the same gives,
\be
g_{\mu \nu} = \eta_{ab}e_\mu^a e_\nu^b~~~\leftrightarrow~~~
g^{\mu \nu} = \eta^{ab}e^\mu_a e^\nu_b
\label{vierbeindef.eqn}
\ee
which defines the frame fields $e_\mu^a$.
In units of $v_F=1$ the tilt metric changes to flat space $\eta_{ab}$ to
\begin{eqnarray}
ds^2  &=& (-1+\zeta^2)dt^2 -2\zeta_x dx dt -2\zeta_y dy dt +dx^2 + dy^2\nn\\
  &=& -dt'^2 + du^2 + dv^2
\end{eqnarray}
which can be obtained by a affecting a {\em Galilean} boost in Minkowski spacetime:
\begin{eqnarray}
&x'^0 = t' = t, \nonumber\\
&x'^1 = u = x-\zeta_x t,\nonumber\\
&x'^2 = v= y-\zeta_y t.
\end{eqnarray}
Using the transformation law of tensors,
\begin{eqnarray}
g_{\mu \nu} = \frac{\partial x'^a}{\partial x^\mu} \frac{\partial x'^b}{\partial x^\nu} \eta_{ab}
\end{eqnarray}
and comparing with Eq.~\eqref{vierbeindef.eqn} gives the frame fields $e_\mu^{~a} = \frac{\partial x'^a}{\partial x^\mu}$ as follows:
\begin{eqnarray}
&e_0^{~0} = \frac{\partial t'}{\partial t} =1, ~~~~
e_0^{~1} = \frac{\partial u}{\partial t} =-\zeta_x, ~~~~
e_0^{~2} = \frac{\partial v}{\partial t} =-\zeta_y, \nonumber \\
&e_1^{~0} = \frac{\partial t'}{\partial x} =0,~~~~
e_1^{~1} = \frac{\partial u}{\partial x} =1,~~~~
e_1^{~2} = \frac{\partial v}{\partial x} =0, \nonumber \\
&e_2^{~0} = \frac{\partial t'}{\partial y} =0,~~~~
e_2^{~1} = \frac{\partial u}{\partial y} =0,~~~~
e_2^{~2} = \frac{\partial v}{\partial y} =1 .
\end{eqnarray}
From these frame fields one can then construct, 
\be
\omega_\mu ^{a b} = e^a_{\lambda}(x)g^{\lambda \sigma}(x)\nabla_\mu e^{b}_{\sigma}(x)
\ee
where the covariant derivative of the frame fields are  
\begin{eqnarray}
\nabla_\mu  e^a_{\sigma} = \partial_\mu e^a_{\sigma} -\Gamma_{\mu \sigma}^{~~\lambda} e^a_{\lambda}.
\end{eqnarray}
\begin{widetext}
To construct spin connection we calculate $\omega^{ab}_\mu$ as follows:
\begin{eqnarray}
& \omega_\mu^{12} = e^1_\lambda g^{\lambda \sigma} \nabla_\mu e^2_\sigma= e^1_0 g^{0 \sigma} \nabla_\mu e^2_\sigma + e^1_1 g^{1 \sigma} \nabla_\mu e^2_\sigma+ e^1_2 g^{2 \sigma} \nabla_\mu e^2_\sigma \nn\\
& =(e^1_0g^{00} + e^1_1 g^{10} + e^1_2 g^{20})\nabla_\mu e^2_0  +(e^1_0g^{01} + e^1_1 g^{11} + e^1_2 g^{21})\nabla_\mu e^2_1 +(e^1_0g^{02} + e^1_1 g^{12} + e^1_2 g^{22})\nabla_\mu e^2_2,
\end{eqnarray}
so that
\begin{eqnarray}
\omega_\mu^{12} 
=(e^1_0g^{00} + e^1_1 g^{10} )\nabla_\mu e^2_0  +(e^1_0g^{01} + e^1_1 g^{11} )\nabla_\mu e^2_1 +(e^1_0g^{02} + e^1_1 g^{12} )\nabla_\mu e^2_2. 
\end{eqnarray}  
Next we have 
\begin{eqnarray}
\nabla_\mu e^2_0=\partial_{\mu}e^2_0-(\Gamma^0_{\mu0}e^2_0+\Gamma^2_{\mu0}),~~
\nabla_\mu e^2_1=-(\Gamma^0_{\mu1}e^2_0+\Gamma^2_{\mu1}) ,~~
\nabla_\mu e^2_2=-(\Gamma^0_{\mu2}e^2_0+\Gamma^2_{\mu2}) 
\end{eqnarray}
and furthermore, 
\begin{eqnarray}
\omega_\mu^{12} =(\zeta_y\Gamma^0_{\mu1}-\Gamma^2_{\mu1})=0.
\end{eqnarray}
In the same way we able to find other components of $\omega_\mu^{ab}$,
\begin{eqnarray}
& \omega_\mu^{02} = e^0_\lambda g^{\lambda \sigma} \nabla_\mu e^2_\sigma = e^0_0 g^{0 \sigma} \nabla_\mu e^2_\sigma + e^0_1 g^{1 \sigma} \nabla_\mu e^2_\sigma+ e^0_2 g^{2 \sigma} \nabla_\mu e^2_\sigma \nonumber\\
& =(e^0_0g^{00})\nabla_\mu e^2_0 
 +(e^0_0g^{01})\nabla_\mu e^2_1 
 +(e^0_0g^{02})\nabla_\mu e^2_2 
=-\nabla_\mu e^2_0 -\zeta_x \nabla_\mu e^2_1 -\zeta_y\nabla_\mu e^2_2 \nonumber\\
&=-\partial_{\mu}e^2_0+(-\zeta_y\Gamma^0_{\mu0}+\Gamma^2_{\mu0}) +\zeta_x(-\zeta_y\Gamma^0_{\mu1}+\Gamma^2_{\mu1}) +\zeta_y(-\zeta_y\Gamma^0_{\mu2}+\Gamma^2_{\mu2}) 
\end{eqnarray}
\begin{eqnarray}
& \omega_\mu^{01} = e^0_\lambda g^{\lambda \sigma} \nabla_\mu e^1_\sigma= e^0_0 g^{0 \sigma} \nabla_\mu e^1_\sigma + e^0_1 g^{1 \sigma} \nabla_\mu e^1_\sigma+ e^0_2 g^{2 \sigma} \nabla_\mu e^1_\sigma \nonumber\\
& =-\nabla_\mu e^1_0 -\zeta_x\nabla_\mu e^1_1 -\zeta_y\nabla_\mu e^1_2 \nonumber\\
& =\partial_{\mu}\zeta_x+(-\zeta_x\Gamma^0_{\mu0}+\Gamma^1_{\mu0}) +\zeta_x(-\zeta_x\Gamma^0_{\mu1}+\Gamma^1_{\mu 1}) +\zeta_y(-\zeta_x\Gamma^0_{\mu2}+\Gamma^1_{\mu2}).
\end{eqnarray}
\end{widetext}
Therefore the non-zero components of the spin connection are eventually given by
\begin{eqnarray}
&\omega_{0}^{02} =- \frac{1}{2}\zeta_x\left( \partial_x\zeta_y +\partial_y\zeta_x \right) - \zeta_y\partial_y\zeta_y \nonumber\nn\\
&\omega_{0}^{01} =- \frac{1}{2}\zeta_y\left( \partial_x\zeta_y +\partial_y\zeta_x \right) - \zeta_x\partial_x\zeta_x \nn\\
&\omega_{x}^{01}= \partial_x \zeta_x \nonumber\\
&\omega_{x}^{02} = \frac{1}{2}\left( \partial_x\zeta_y +\partial_y\zeta_x \right) \nonumber\\
&\omega_{y}^{01} =\frac{1}{2}\left( \partial_x\zeta_y +\partial_y\zeta_x \right) \nonumber\\
&\omega_{y}^{02} = \partial_y \zeta_y 
\label{spinconnexpanded.eqn}
\end{eqnarray}
If one defines following quantity, 
\begin{eqnarray} 
\alpha_i^j= \frac{1}{2} \left(\partial_i \zeta^j + \partial_j \zeta^i \right)
\end{eqnarray}
the compact form of above equations result to, 
\bearr
\omega_i^{~0j}=\alpha_i^{~j},~\omega_0^{~0i}=-\alpha_i^{~j}\zeta_j,~\mbox{others}=0. 
\eearr
Now that we have all the components $\omega_\mu^{ab}$
we can contract it with the generators $\Sigma_{ab}=[\gamma_a,\gamma_b]/4$ of the Lorentz group to obtain the spin connection
$  \Omega_\mu = \frac{1}{2}\omega_\mu ^{a b}\Sigma_{ab}$. 
As we discussed in the main text of the paper, definition of covariant derivative in the presence of 
variable $e^\mu_a$ (variable frame fields) is, 
\begin{eqnarray}
\nabla_a =e^\mu_a \ (\partial_\mu+ \Omega_\mu)
\end{eqnarray}
when the frame fields are not position dependent, the above equation reduce to, 
\begin{eqnarray}
\nabla_a = e^\mu_a \partial_\mu
\end{eqnarray}
or
\begin{eqnarray}
\left(\begin{matrix}
\nabla_0 \\
\nabla_1 \\
\nabla_2 
\end{matrix}\right) = \left( \begin{matrix}
1 & \zeta_x & \zeta_y \\
0 & 1 & 0 \\
0 & 0 & 1
\end{matrix}\right) \left(\begin{matrix}
\partial_0 \\
\partial_1 \\
\partial_2 
\end{matrix}\right)
\end{eqnarray}
so that the Lagrangian density for a tilted system is
\begin{eqnarray}
\mathcal{L} = i \bar{\psi} \left( \gamma_\mu \partial_\mu + \gamma_0 \vec\zeta \cdot \vec \partial \right) \psi + c\cdot c 
\end{eqnarray}
but for a variable (position dependent) tilt we get the gauge fields as, 
\begin{eqnarray}
\mathcal{A}_a =  e^\mu_a \Omega_\mu 
\end{eqnarray}
This equation establishes that the space dependence of the tilt parameters $\bs\zeta$ in metric~\eqref{metric.eqn},
induces non-Abelian gauge fields given by Eq.~\eqref{gauges.eqn}.

\section*{References}
\bibliographystyle{apsrev4-1}
\bibliography{Refs}
\end{document}